# DEGRADATION OF NONYLPHENOL ETHOXYLATE-10 (NPE-10) BY MEDIATED ELECTROCHEMICAL OXIDATION (MEO) TECHNOLOGY

Henry Setiyanto[1, 2,] *, Muhammad. Muslim Syaifullah[1], I Made Adyatmika[1], Dian Ayu Setyorini[1], Muhammad Yudhistira Azis[1], Vienna Saraswaty[3], Muhammad Ali Zulfikar[1]

*1Analytical Chemistry Research Group, Institut Teknologi Bandung, Bandung, Indonesia*

*2Center for Defense and Security Research, Institut Teknologi Bandung, Bandung, Indonesia*

*3Research Unit for Clean Technology, Indonesia Institute of Sciences, Bandung, Indonesia*

*Corresponding author: henry@chem.itb.ac.id*

**Abstract**

Nonylphenol ethoxylate (NPE–10) is a non-ionic surfactant which is synthesized from alkylphenol ethoxylate. The accumulation of NPE-10 in wastewater will endanger the ecosystem as well as human being. At present, by an advancement of technology NPE-10 can be degraded indirectly by using an electrochemically treatment. Thus, this study aimed to evaluate the potential electro-degradation of NPE-10 by mediated electrochemical oxidation (MEO) using Ce(IV) ionic mediator. In addition, the influence of Ag(I) ionic catalyst in the performance of MEO for degradation of NPE-10 was also observed. The potency of MEO technology in degradation NPE-10 was evaluated by voltammetry technique and confirmed by titrimetry and LC-MS analyses. The results showed that in the absence of Ag(I) ionic catalyst, the degradation of NPE-10 by MEO was 85.93 %. Furthermore, when the Ag(I) ionic catalyst was applied, the performance of MEO in degradation of NPE-10 was improved to 95.12 %. The back titration using Ba(OH)$_2$ confirmed the formation of $CO_2$ by 46.79 %. Whereas the redox titration shows the total of degradation organic compounds by 42.50 %, which was emphasized by formation of two new peaks in LC-MS chromatogram. In summary, our results confirm the potential of MEO technology for NPE-10 degradation.

**Introduction**

Nonylphenol ethoxylate (NPE) is one of non-ionic surfactant in the class of alkylphenol ethoxylates. NPE is widely used as dispersing agent in household, industrial applications, detergent, as well as emulsifier (Martins et al., 2006; Fuente et al., 2010; Brigden et al., 2012). The toxicity of NPE is relatively low, however concerning to the increment of the widespread usage of this organic pollutant and its potential to produce highly toxic stable metabolites such as ethylene oxide, as well as its slow biodegradation, NPE can cause serious environmental problems. More importantly, according to toxicity evaluation, this compound also considered as an endocrine disrupting compound which may result in functional abnormalities as well as cancer (Liu et al., 2017; Forte et al., 2016; Shufaro et al., 2018). Several approaches have been developed for NPE degradation including biological treatment, as shown by studies of Lu et al., (2007), Maki et al. (1994) and Mao et al. (2012), Hernandez-Raquet et al. (2007) whose results successfully degraded NPE in ranging of 60-100%. However, those biological treatments of NPE may cause other problems as their potential in reducing oxygen transfer, producing foams, disturbing the sedimentation process, as well as disturbing the ecosystem. For these reasons, procedures in degradation of NPE need to be established.

Electrochemical based technology offers advantages including environmental compatibility, versatile and cost effective. Mediated Electrochemical Oxidation (MEO) is one of electrochemical treatments which commonly used for degradation of organic pollutants based on the activity of oxidant species (ionic mediators), including Ce(IV), Cl· and $S_2O_8^{2-}$ (Juttner et al., 2000). Due to the nature of NPE as an electro-active organic compound, MEO is possible to be applied for degradation of NPE. The principle of redox reaction is by electron transfer mechanism (Setiyanto et al. 2011; Seo et.al, 2017; Jamshidinia et. al, 2017; Setiyanto et. al, 2011; Cha et.al, 2017; Cho et.al, 2017; Setiyanto et. al, 2015) was used in this study.

NPE-10 (Figure 1) is one of the most common nonionic surfactant (Cox et al., 1984, Olkowska et al., 2014). The maximum concentration of NPE–10 allowed in the freshwater and saltwater should not exceed 6.6 µg/L and 1.7 µg/L (David et al., 2009; Brooke & Thursby, 2005). Since no study has been reported about degradation of NPE-10 by MEO technology, we therefore interested to evaluate the potential of MEO in degradation of NPE-10.

Figure 1. Structure of Nonylphenol Ethoxylates-10 (NPE-10)

In this paper, the potential of MEO in degradation of NPE-10 was evaluated by voltammetry techniques and confirmed by back and redox titration, and LC-MS analyses. In addition, we also evaluated the influence of addition of $Ag^+$ ionic catalyst in the MEO performance for NPE-10 degradation.

**Materials and Methods**

**Apparatus**

The degradation process was carried out by using Potensiostat/Galvanostat eDAQ 410 with Pt wire as the cathode and anode. The degradation product was evaluated by voltammetric techniques using BASi Epsilon Electrochemical Analyzer. The voltammetric detection used three electrodes, i.e. carbon paste electrode (CPE) as the working electrode, Ag/AgCl as the reference electrode, and Pt wire as the auxiliary electrode. The product of degradation was further analyzed using LC-MS XEVO – Qtof – MS instrument.

## Chemicals

The cerium sulfate octahydrate (> 99.0%) was purchased from (Sigma Aldrich). Sodium chloride, barium dioxide, potassium permanganate, oxalate acid, hydrochloric acid, $Al_2O_3$, sulphuric acid and graphite were from eMerck with pro analytical grade. Whereas NPE-10 was purchased from a local textile shop with technical grade.

## Preparation of the working electrode

The working electrode (CPE) was prepared by heating the mixture of carbon and liquid paraffin (in the ratio of 7:3) at 80°C in a beaker glass. The mixtures were then inserted into a cylindrical electrode holder and cooled in room temperature. For the electrical contact, the carbon paste was connected with a copper wire. The surface of the prepared working electrode was then polished by alumina ($Al_2O_3$) slurry onto a polishing pad.

## Sample preparations and voltammetric measurement

The oxidation potential of Ce(III) and NPE-10 were investigated by using voltammetric method. Four solutions were prepared in 0.2 M $H_2SO_4$: (1) background solution (0.4 M $H_2SO_4$); (2) 0.04 M Ce (III); (3) 1000 ppm NPE-10; (4) the mixture of 0.04 M Ce(III) and 1000 ppm NPE-10. The measurement was recorded in the scan range of -200 - 2000 mV with the scan rate of 100 mV/s using cyclic voltammetry.

For measurement optimization, four solutions were prepared in 0.4 M $H_2SO_4$ containing: (a)

1500 ppm NPE-10; (b) 0.04 M Ce(III) and 1500 ppm NPE-10; (c) 0.04 M Ce(III), 0.009 M Ag(I), and 1500 ppm NPE-10. The duration of degradation was varied from 5 to 55 minutes. The solution after degradation was further evaluated by voltammetric techniques.

*LC-MS Analysis*

LC-MS analysis was performed using LC-MS XEVO – Qtof - MS instrument. The column used was ACQUITY UPLC BEH C18 (1,7 μm x 2,1 mm x 100 mm). The flow rate was 0.3 mL/min with injection volume of 5 μL. For the separation of degradation products, the following solvents were used: 2 mM ammonium acetate in water (A) and 2 mM ammonium acetate in acetonitrile (B).

**Indirect determination of carbon dioxide by back titration.**

The content of $CO_2$ gas formed during degradation of NPE-10 by MEO was evaluated by back titration using $Ba(OH)_2$ solution. The amount of $Ba(OH)_2$ is equivalent with $CO_2$ according the following reaction :

$$Ba(OH)_2(aq) + CO_2(g) \rightarrow BaCO_3(s) + H_2O(l) \quad (1)$$

The protocol for indirect determination of $CO_2$ is as follow: the $CO_2$ gas formed during degradation of NPE-10 by MEO was fed into an Erlenmeyer flask containing $Ba(OH)_2$ solution ($Ba(OH)_2$ total at 0.9588 mmol). After degradation process finished, the $Ba(OH)_2$ solution containing $CO_2$ was then titrated with 0.0282 mmol HCl. The mmol of $Ba(OH)_2$ remained was assumed as the amount of $Ba(OH)_2$ that was not reacted with the $CO_2$ formed from NPE-10 degradation.

$[Ba(OH)_2]$ (mmol) = $[Ba(OH)_2]$ total (mmol) – $[Ba(OH)_2]$ titration (mmol).

The percentage of $CO_2$ formed can be calculated by the following formula

$$\% \, CO_2 = \frac{mmol \; of \; Ba(OH)2}{mmol \; CO2 \; theoretical} \times 100\%$$

Where mmol $CO_2$ theoretical according to the prediction of mmol $CO_2$ from the propose NPE-10 complete degradation reaction on anode (Balaji et al., 2007; Chung & Park, 2000).

$$C_{35}H_{64}O_{11}/(NPE-10) + 182Ce(IV) + 59H_2O \rightarrow 35CO_2 + 182Ce(III) + 182H^+ + 182e^- \quad (2)$$

**Determination of permanganate value by redox titration.**

The percentage of degraded organic compound from NPE-10 degradation by MEO was confirmed by redox titration. Prior redox titration, the samples were prepared as follow: The solution of NPE-10 before and after degradation were diluted for 25 times. About 100 mL of NPE-10 solution was put into a 300 mL Erlenmeyer flask and added with a few drops of 0.01 N $KMnO_4$ until the solution color becomes pink. The solution was further added with 5 mL of 8 N sulfuric acid and heated to 105 °C then colled in room temperature. About 10 mL of prepared sample was then pipetted and transferred into an Erlenmeyer flask. Subsequently, the sample was re-heated for 10 minutes. After 10 minutes, the solution was added with 10 mL of 0.01 N oxalic acid. The excess oxalic acid was titrated with 0.01 N $KMnO_4$. The permanganate value was calculated by using the following formula:

$$KMnO_4 (mg/L) = \frac{[(10-a) \times b - (10 \times c)] \times 1 \times 31.6 \times 1000}{d} \times f$$

Where is, "a" is the required volume of $KMnO_4$ in the titration, "b" is the actual $KMnO_4$ normality, "c" is the normality of oxalic acid, "d" is the sample volume, and "f" is the dilution factor.

The percentage degraded organic compound =

$$\frac{[KMnO4\ value]\ before\ degradation - [KMnO4\ value]\ after\ degradation}{[KMnO4\ value]\ before\ degradation} \times 100\%$$

(Suslova et al., 2014)

**Results and Discussion**

## Voltammetric study of Ce(III)/Ce(IV) and NPE-10

The voltammetric behavior of Ce(III) and NPE-10 were observed by using cyclic voltammetry (CV) in $H_2SO_4$ as the supporting electrolyte (Paulenova et al., 2002; Ren and Wei, 2011). The measurement of CV was aimed to determine whether the reaction is reversible or irreversible. As shown in Figure 2, two reduction peaks were observed at potential value of 0.786 V and 0.444 V, respectively. Those peaks were assumed as reduction peaks of $H_2SO_4$, as suggested by the following prediction reduction reaction:

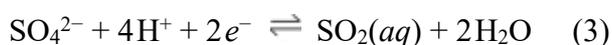
$$SO_4^{2-} + 4H^+ + 2e^- \rightleftharpoons SO_2(aq) + 2H_2O \quad (3)$$

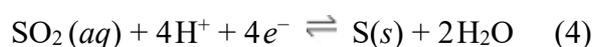
$$SO_2(aq) + 4H^+ + 4e^- \rightleftharpoons S(s) + 2H_2O \quad (4)$$

As also can be seen in Figure 2, the voltammogram of Ce(III)/Ce(IV) (black line) showed oxidation and reduction peak current (Ipa and Ipc) at 1.322 V and 1.148 V repectively, showing reversible reaction. Meanwhile, the voltammogram of NPE-10 (red line) only showed oxidation peak current at 1.400 V, indicating irreversible reaction of NPE-10. As also depicted in Figure 2 (blue line), in the mixture of Ce(III) and NPE-10 solution, the nett of NPE-10 peak oxidation current (ΔIp) decreased, indicating the oxidation of NPE-10 due to the reduction reaction of Ce(IV) becomes Ce(III).

Figure 2. The cyclic voltammogram of (a) 0.04 M Ce(III), (b) NPE-10, (c) 0.04 M Ce(III) and NPE-10 in 0.4 M $H_2SO_4$ by CPE as the working electrode, Ag/AgCl as the reference electrode and Pt wire as the auxilary electrode.

## Selection of voltammetry technique for evaluation of NPE-10 degradation

Prior to the evaluation of NPE-10 degradation, we selected the most suitable voltammetry techniques as suggested by the highest nett current. In this study four voltammetry techniques were evaluated for NPE-10 degradation that are linear sweep voltammetry (LSV), cyclic voltammetry (CV), differential pulse voltammetry (DPV), and square wave voltammetry (SWV). The NPE-10 solution used in this study was prepared at 5000 ppm in 0.4 M $H_2SO_4$.

The value of nett current was presented in Figure 3.

Figure 3. The nett current of NPE-10 using various voltammetry technique.

As depicted in Figure 3, the nett current values of NPE-10 were 0.0477 mA; 0.0491 mA; 0.0777 mA and 0.0990 mA for LSV, CV, DPV and SWV, respectively. Comparison between nett current value of NPE-10 using SWV and other techniques showed that SWV more sensitive than that of other techniques. It has been reported that the limit detection of DPV and SWV were in nanomolar range which are superior than LSV and CV (Hussain & Silvester, 2018). In addition, in DPV technique, only forward current was measured. Whereas in SWV technique both the forward and reverse current are measured, hence the higher nett current value was observed by SWV. Our results were in agreement Wang (2000), which reported that SWV was 3.3 times more sensitive than DPV for detection of irreversible compounds.

**The optimization of NPE-10 degradation time**

In waste treatment, degradation time plays important role particularly in cost effectiveness. In general, the longer degradation time, the more compound or organic pollutants are degraded and the higher cost is required. However, when the steady state condition is reached, no more compounds will be degraded. Thus, it is very important to optimize the degradation time NPE-10. In our previous report, we successfully electro-degraded NPE-10 by MEO at potential value of 5.5 V (Muslim et al. 2018). The percentage degradation value of NPE-10 was ~80%. In this study, the percentage degradation of NPE-10 as function of time was presented in Figure 4. As shown, the percentage degradation of NPE-10 tends to increase at minute of 5 to 35. Whereas at minute of 35 to 55, the percentage degradation of NPE-10 is relatively constant. The percentage degradation value of NPE-10 at minute of 35 was 85.51%. Accordingly, we suggest applying 35 minutes of degradation time for degradation of 1500 ppm NPE-10 by MEO.

Figure 4. Percentage degradation of NPE-10 by MEO as function of time.

**The effect of Ag(II) catalyst ion for NPE-10 degradation by MEO using Ce(IV) mediator ion**

The influence of Ag(I) ionic catalyst addition in the NPE-10 degradation by MEO was presented in Table 1. As presented, the degradation of NPE-10 by MEO in acidic solution of without Ce(III), with Ce(III) and combination of Ce(III):Ag(I) were 61.44%, 85.93% and 95.12% respectively. From Table 1, it is obviously seen that, the presence of Ce(III) and Ag(I) increased the NPE-10 degradation by MEO. It should be noted that in an acidic solution containing Ce(III), the Ce(IV) can be produced as the result of redox reaction. The produced Ce(IV) will continuously oxidize NPE-10 (See Figure 5) hence resulting degradation products including $CO_2$ (Raju & Basha, 2005; Martinez-Huitle & Ferro, 2006).

**Figure 5.** The degradation mechanism of NPE-10 by MEO using Ce(III) as ionic mediator.

In an acidic solution, Ag(I) will be oxidized as Ag(II). Ag(II) is an unstable ion and can act as an oxidator. Therefore, the presence of Ag(II) in solution containing Ce(III) supports the oxidation of Ce(III) to Ce(IV). According the results of NPE-10 degradation in the presence of Ag(I), we revealed that the performance of NPE-10 by MEO was improved. Furthermore, our data were in the line with previous studies which have reported that the presence of Ag(I) in a solution containing Ce(III) will increase the oxidation peak current of Ce(III) (Fawzy & Al-Jahdali 2016; Sumathi et al. 2010; Matheswaran et al. 2007). Based on the explanation above, Ag(I) has potential as a ionic catalyst in the redox reaction of Ce(III)/Ce(IV). Therefore, we suggest to add Ag(I) in acidic solution of Ce(III) for the electro-degradation of NPE-10 by MEO for a better result.

Table 1. The percentage degradation of NPE-10 in acidic solution

**Degradation of NPE-10 by MEO resulting $CO_2$**

In anode, the degradation of NPE-10 was predicted forming $CO_2$ (See the propose reaction 2). We confirmed the formation of $CO_2$ from NPE-10 degradation according to precipitation of white sendiment as the results of reaction $Ba(OH)_2$ with $CO_2$. This result was also emphasized by calculation of $CO_2$ formed by back titration. It shows that NPE-10 resulted in 46.79% $CO_2$ formation. The results from back titration obviously showed the incompleted degradation of NPE-10 by MEO. It should be noted that results from degradation of organic pollutants are not always in the form of $CO_2$ as their final product. An organic pollutant may also produce more simple compounds as its degradation product (Brillas, 2014, Setiyanto et al., 2016). In the respect to the result of $CO_2$ analysis concentration, we therefore suggest that the remaining content were other organic compounds with smaller molecules.

**Characterization of degradation products using LC – MS.**

Results from Table 1 obviously showed the potential of MEO for electro-degradation of NPE-10 and in producing smaller molecules. For this reason, we further analysis the degradation products of NPE-10 by LC-MS analysis. The LC-MS spectrum of NPE-10 before and after degradation is presented in Figure 6. As can be seen, before degradation process, the NPE-10 shows peaks at retention time of 5.00; 5.41; 5.75; 9.72; 11.07; and 15.05 minute. This due to the NPE-10 used in this study is a technical grade, thus showing its impurities. However, the impurities of NPE-10 used in this study did not interfere. The degradation products were detected in Figure 6, as suggested by two new peaks present in the LC-MS chromatogram of NPE-10 after degradation process (Figure 6) at retention time of 6.970 and 13.618 minute. The ion chromatograms and characterization of those compounds can be seen in Figure 7 and 8. As shown in Figure 7, for compound at retention time of 6.970 minute, the most significant ion is produced at m/z of 413.29, corresponding to the combination of benzylic cleavages $[M-71]^+$ and the shortened of polyethoxy chain $[M-176]^+$. Whereas in Figure 8, for compound at retention time of 13.618 min, the most significant ion is produced at m/z of 603.92, corresponding to benzylic cleavage of $[M-57]^+$. The benzylic cleavages and the shortened of polyethoxy chains are common for degradation of NPE compounds (Karci, 2014; Namara et al., 2012; Li et al., 2018). According to the ion at m/z of 603.92, it seems that the terminal atom carbon of hydrophobic part of NPE-10 was firstly oxidized. And subsequently, the polyethoxylate chain which was continuously oxidized. Hence, resulting smaller molecules. Although the ion chromatograms in Figure 7 and 8 are not very informative in respect with structure elucidation, however, they provide an indication of degradation products NPE-10 and create a landscape for the future development of NPE-10 degradation process.

Figure 6. LC-MS chromatogram of NPE-10 before and after degradation by MEO.

Figure 7. Ion chromatogram of degradation compound from NPE-10
at retention time of 6.97 min

Figure 8. Ion chromatogram of degradation compound from NPE-10
at retention time of 13.61 min

**Determination of organic matter content using permanganate value**

The permanganate value denotes the number of milligrams of $KMnO_4$ required in oxidation of organic substances as suggested by the following reaction:

$$C + MnO_4^- + 4H^+ + e \rightarrow CO_2 + Mn^{2+} + 2H_2O \quad (5)$$

(Suslova et al., 2014). Our results show that before and after degradation of NPE-10 the KMnO$_4$ value were 445.56 mg/L and 256.18 mg/L, respectively. The percentage degradation organic compound was determined as 42.50 %. By summarizing the percentage value of CO$_2$ and degradation organic compound, about 88% of NPE-10 was assumed degraded. This result almost similar with calculation of degradation of NPE-10 by voltammetry technique as suggested by degradation of 95.1% NPE-10. It is not surprisingly that the result of degradation of NPE-10 by MEO using voltammetry technique is higher than the value resulted from back and redox titration. Because in voltammetry technique, the value recorded based on the reaction on the electrode surface. Meanwhile in the back and redox titration, the values obtained by using bulk solution. In addition, the more complex reaction may be occurred in a bulk solution, hence resulting in lower value. In reverse, on the surface of electrode, the more simple and certain reaction occur as proposed in Figure. 5, hence resulting in higher value. The results from voltammetry technique and titration are mutually support each other.

## Conclusion

In conclusion, NPE-10 can be degraded by using MEO technology. About 88-95% of 1500 ppm of NPE-10 can be degraded by MEO within 35 minutes. The presence of Ag(I) ionic catalyst improves the performance of MEO technology for degradation of NPE-10. Thus, we suggest that the MEO technology is a promising method for degradation of organic pollutant in waste water.

## Acknowledgements

This research has been partially supported by Analytical Chemistry Research Group, Institut Technologi Bandung (Indonesia) and Ministry of Research and Higher Education of Indonesia through Riset PUPT 2017-2018. Authors would like to express gratitude to M.A. Majid Al Kindi for proofreading this manuscript.

# References

Balaji, S., Sang, J.C., Manickam, M., Kokovkin, V. V., & Moon, I.S. 2008. Destruction of Organic pollutants by cerium(IV) MEO process: A study on the influence of process conditions for EDTA mineralization. *Journal of Hazardous Materials* 150: 596-603.

Brillas, E. 2014. A review on the degradation of organic pollutants in waters by UV photoelectro-Fenton and solar photoelectro-Fenton. *Journal of Brazillian Chemical Society* 25: 393 – 417.

Brigden, K., Santillo, D. & Johnston, P. 2012. Nonylphenol ethoxylates (NPEs) in textile products, and their release through laundering. *Greenpeace Research Laboratories Technical Report.*

Brooke, L. & Thursby, G. 2005. Ambient aquatic life water quality criteria for nonylphenol. *Report for the United States EPA, Office of Water, Office of Science and Technology, Washington, DC, USA.*

Chung, Y.H. & Park, S.M. 2000. Destruction of anilin by mediated electrochemical oxidation with Ce(IV) and Co(III) as mediators. *Journal of Applied Electrochemistry* 30: 685 – 691.

Cox, M.F. & Matson, T.P. 1984. Optimization of nonionic surfactants for hard-surfaces cleaning. *Journal of the American Oil Chemist Society* 61: 1273 – 1278.

David, A., Helene F. & Elena G. 2009. Alkylphenols in marine environments: Distribution monitoring strategies and detection considerations. *Marine Pollution Bulletin* 58: 953 – 960.

Fawzy, A. & Al – Jahdali B.A. 2016. Silver(I) Catalysis for Oxidation of L-Glutamine By Cerium(IV) in Perchlorate Solutions: Kinetics and Mechanistic Approach. *Journal of Austin Chemical Engineering* 3(4): 1037.

Forte, M., Lorenzo, M.D., Zarrizzo, A., Valiante, S., Vecchione, C., Laforgia, V., & Falco, M.D.

2016. Nonylphenol effects on human prostate non tumorigenic cells. *Toxicology* 357-358: 21-32.

Fuente. L.D.L., Acosta, T., Babay. P., Curutchet, G., Candal, R. & Litter. M.I. 2010. Degradation of Nonylphenol Ethoxylate-9 (NPE-9) by Photochemical Advanced Oxidation Technologies. *Industrial and Engineering Chemistry Research* 49(15): 6909–6915.

Hernandez-Raquet, G., Soef, A., Delgenès, N., & Balaguer, P. 2007. Removal of the endocrine disrupter nonylphenol and its estrogenic activity in sludge treatment processes. Water Research 41: 2643-2651.

Hussain, G. & Silvester, D.S. 2018. Comparison for voltammetric techniques for amina sensing in ionic liquids. *Electroanalysis* 30: 75 – 83.

Juttner, K., Galla, U. & Schmieder, H. 2000. Electrochemical approaches to environmental problems in the process industry. *Electrochimica Acta* 45: 2575 – 2594.

Karci, A. 2014. Degradation of chlorophenols and alkylphenol ethoxylates, two representative textile chemicals, in water by advanced oxidation process: The state of the art on transformation product and toxicity. *Journal of Chemosphere* 99**:** 1 – 18.

Li, C., Jin, F., & Snyder, S.A. 2018. Recent advancements and future trends in analysis of nonylphenol ethoxylates and their degradation product nonylphenol in food and environment. *Trends in Analytical Chemistry* 107: 78-90.

Liu, C., Lai, Y., Ouyang, J., Yang, T., Guo, Y., Yang, J., & Huang, S. 2017. Influence of nonylphenol and octylphenol exposure on 5-HT, 5-HT transporter, and 5-HT$_{2A}$ receptor. *Environmental Science and Pollution Research* 24(9): 8279-8286.

Lu, J., Jin, Q., He, Y. & Wu, J. 2007. Biodegradation of nonylphenol polyethoxylates under Fe(III)-reducing conditions. *Chemosphere* 69: 1047-1054.

Maki, H., Masuda, N., Fujiwara, Y., Ike, M., & Fujita, M. 1994. Degradation of alkylphenol

ethoxylates by *Pseudomonas* sp. Strain TR01. *Applied and Environmental Microbiology* 60: 2265-2271.

Mao, Z., Zheng, X.F., Zhang, Y.Q., Tao, X.X., Li, Y., & Wang, W. 2012. Occurrence and biodegradation of nonylphenol in the environment. *International Journal of Molecular Sciences* 13(1): 491-505.

Martinez-Huitle, C.A. & Ferro, S. 2006. Electrochemical oxidation of organic pollutants for the wastewater treatment: direct and indirect processes. *Chemical Society Reviews* 35: 1324 – 1340.

Martins, A.F., Wilde, M.L., Vasconcelos, T.G. & Henriques, D. M. 2006. Nonylphenol polyethoxylate degradation by means of electrocoagulation and electrochemical Fenton. *Separation and Purification Technology*, 50:249-255.

Matheswaran, M., Subramanian B., Saan J.C. & Il S.M. 2007. Silver ion catalyzed cerium(IV) mediated electrochemical oxidation of phenol in nitric acid medium. *Journal of Electrochimica Acta* 53: 1897 – 1901.

Muslim, M.S., Setiyanto, H. & Zulfikar, M.A. 2018. Electrodegradation of nonylphenol ethoxylate (NPE-10) with silver ion catalyzed cerium (IV) in sulfuric acid medium. *The Proceedings Book of The 8th Annual Basic Science International Conference 2018*. Page 90.

Namara, P.J.M., Wilson, C.A., Wogen, M.T., Murthy, S.N., Novak, J.T. & Novak. P.J. 2012. The effect of thermal hydrolysis pretreatment on the anaerobic degradation of nonylphenol and short-chain nonylphenol ethoxylates in digested biosolids. *Water Research* 46: 2937 – 2946.

Olkowska, E., Ruman, M. & Polkowska, Z. 2014. Occurrence in surface active agents in the environment. *Journal of Analytical Methods in Chemistry* 769708: 1 – 15.

Paulenova, A., Creager, S. E., Navratil, J.D., & Wei, Y. 2002. Redox potentials and kinetics of the Ce$^{3+}$/Ce$^{4+}$ redox reaction and solubility of cerium sulfates in sulfuric acid. *Journal of Power Sources* 109: 431-438.

Raju, T. & Basha, C.A. 2005. Electrochemical cell design and development for mediated electrochemical oxidation – Ce(III)/Ce(IV) system. *Chemical Engineering Journal* 114: 55 – 65.

Ren, X. & Wei. Q. 2011. A simple modeling study of the Ce(IV) regeneration in sulfuric acids solutions. *Journal of Hazardous Materials* 192: 779-785.

Setiyanto, H., Agustina, D., Zulfikar, M.A. & Saraswaty, V. 2016. Study on the Fenton reaction for degradation of Remazol Red B in textile waste industry. *Molekul* 11(2): 168 – 179.

Setiyanto, H., Saraswaty, V., Hertadi, R., Noviandri, I. & Buchari, B. 2011. Chemical reactivity of chlorambucil in organic solvents ; influence of 4 – chloro butyronitrile nucleophile to voltammogram profile. *International Journal of Electrochemical Science* 6: 2090 – 2100.

Setiyanto, H., Saraswaty, V., Hertadi, R., Noviandri, I. & Buchari, B. 2011. Cyclic voltammetric study of chlorambucil in the presence of 4-chloro butyronitrile in aqueous solution. *International Journal of Chemtech. Research* 3(4): 1986 – 1992.

Setiyanto, H., Saraswaty, V., Hertadi, R., Noviandri, I. & Buchari, B. 2015. Determination of the reactivity of the anti – cancer nitrogen mustard – mechlorethamine : A cyclic voltammetric investigation. *Analytical and Bioanalytical Electrochemistry* 6: 657 – 665.

Shufaro, Y., Saada A., Simeonov, M., Tsuberi, B.Z., Alban, C., Levin, A.K., Shochat, T., Fisch, B., & Abir, R. 2018. The influence of in vivo exposure to nonylphenol ethoxylate 10 (NP-10) on the ovarian reserve in a mouse model. *Reproductive Toxicology* 81: 246-252.

Sumathi, T., Sundaram, P. S. & Chandramohan, G. 2010. A kinetic and mechanistic study on the silver (I) catalyzed oxidation of L-alanine by cerium (IV) in sulfuric acid medium. *Arabian Journal of Chemistry* 4: 427 – 435.

Suslova, O., Govorukha, V., Brovarskaya, O., Matveeva, N., Tashyreva, H., & Tashyerev, O.

2014. Method for determining organic compound concentration in biological systems by permanganate redox titration. *International Journal Bioautomotion* 18(1): 45-52.

Wang, J. 2000. *Analytical electrochemistry*. 2nd Edition. New York: John Wiley&Sons.

**List of figures**

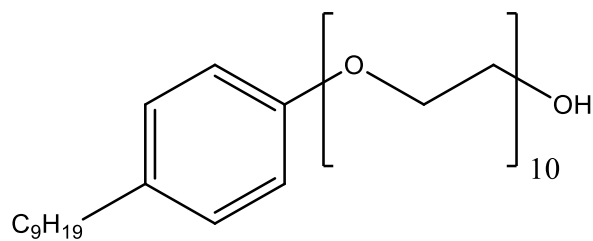

Figure 1. Structure of Nonylphenol Ethoxylates-10 (NPE-10)

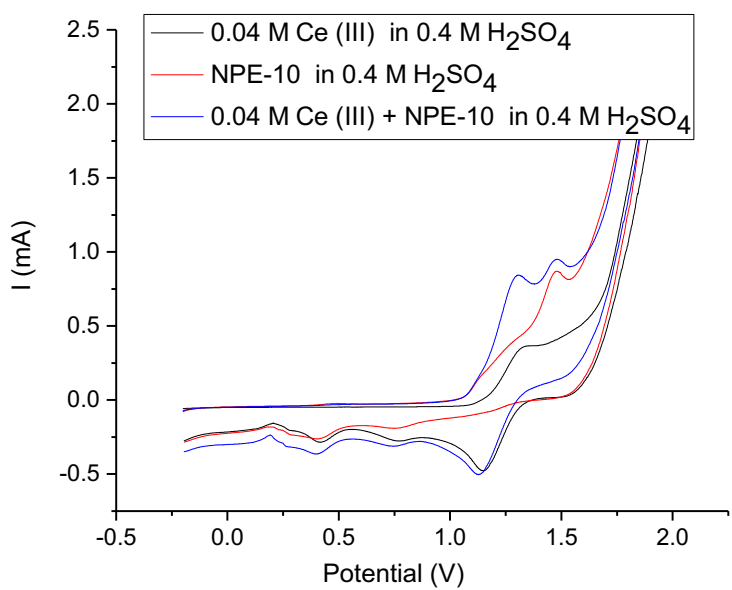

Figure 2. The cyclic voltammogram of (a) 0.04 M Ce(III), (b) NPE-10, (c) 0.04 M Ce(III) and NPE-10 in 0.4 M $H_2SO_4$ by CPE as the working electrode, Ag/AgCl as the reference electrode and Pt wire as the auxilarry electrode.

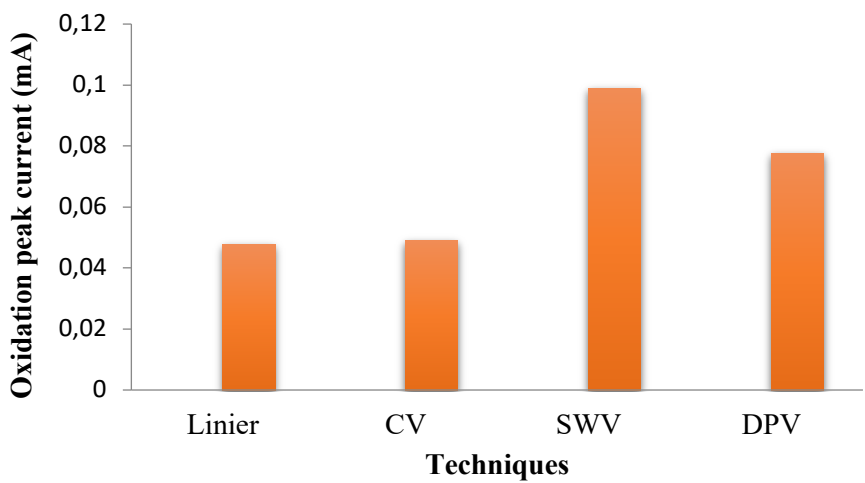

Figure 3. The nett current of NPE-10 using various voltammetry technique.

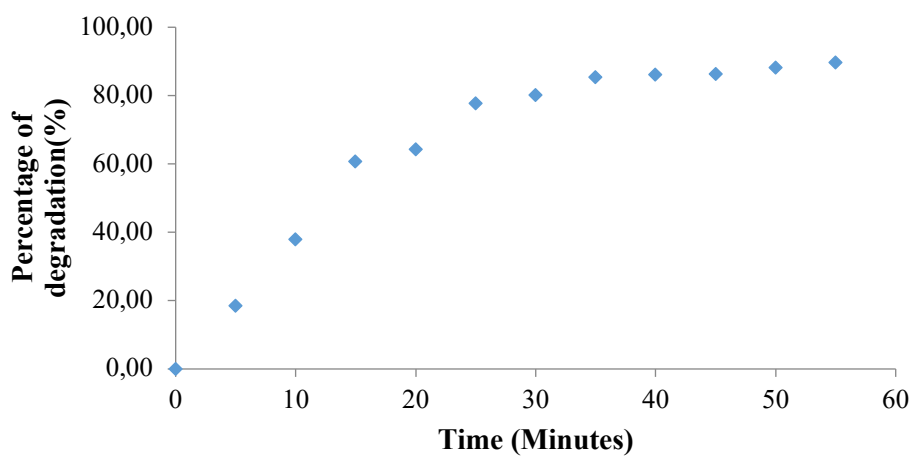

Figure 4. Percentage degradation of NPE-10 by MEO as function of time.

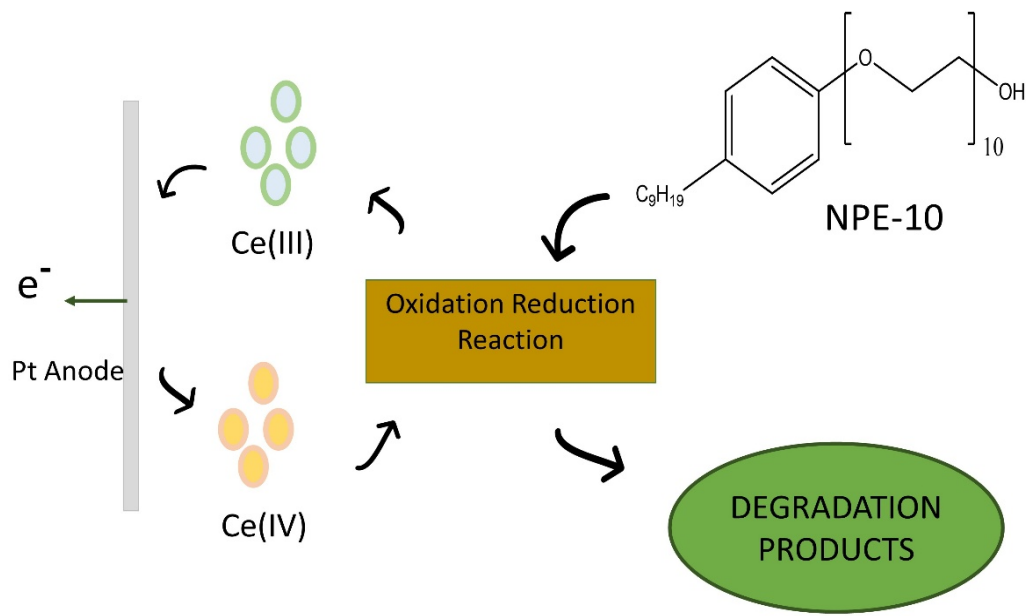

Figure 5. The mechanism process of NPE-10 degradation by using Ce(IV) as mediator ion.

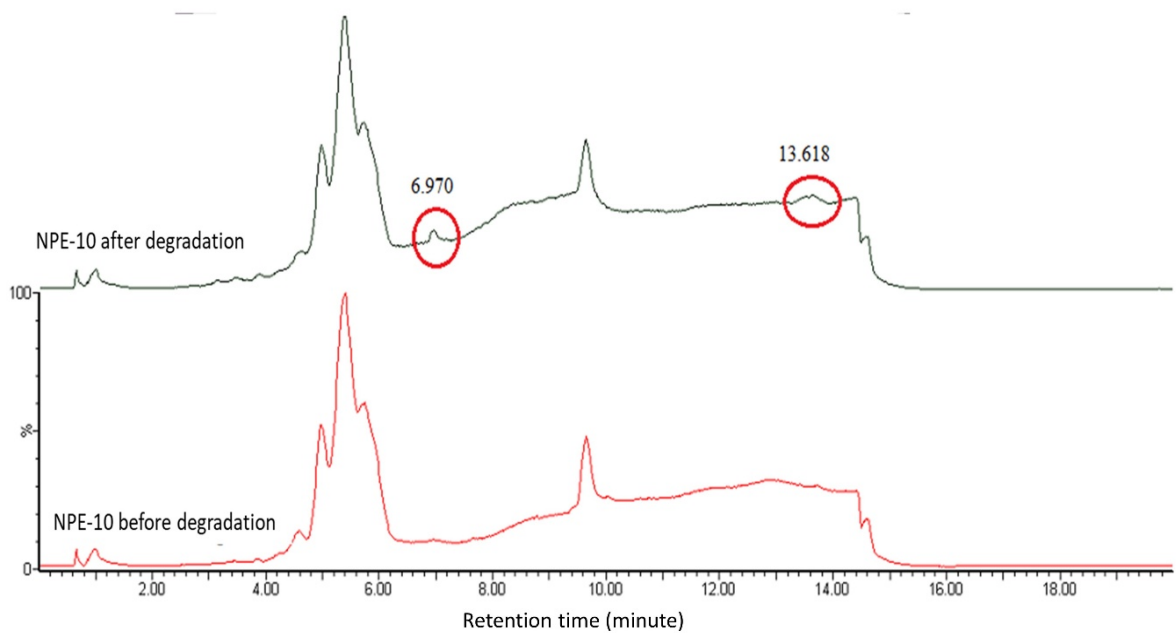

Figure 6. LC-MS chromatogram of NPE-10 before and after degradation by MEO.

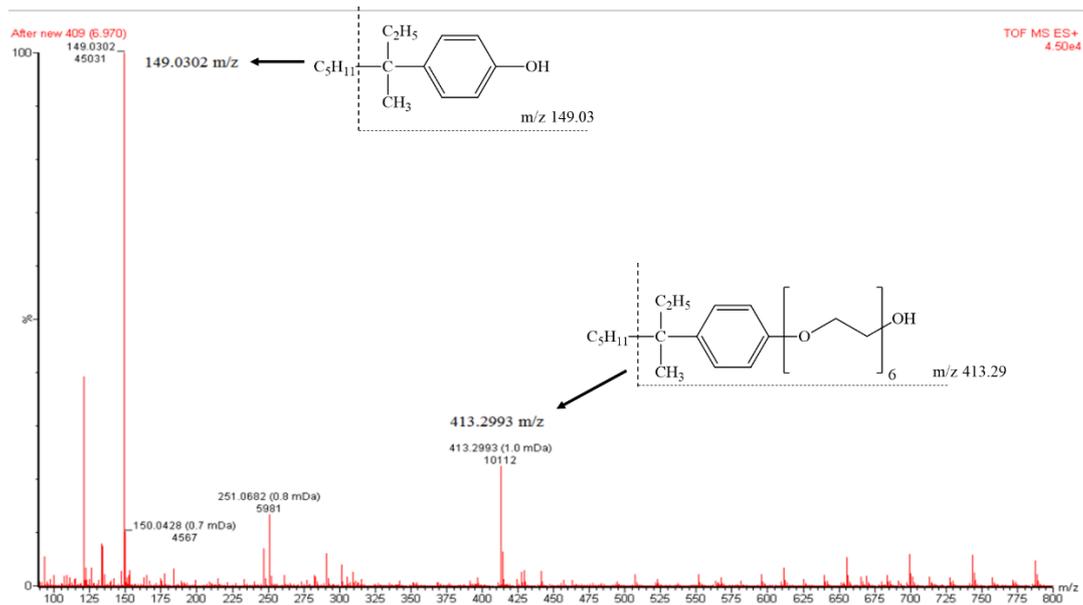

Figure 7. Ion chromatogram of degradation compound from NPE-10 at retention time of 6.97 min.

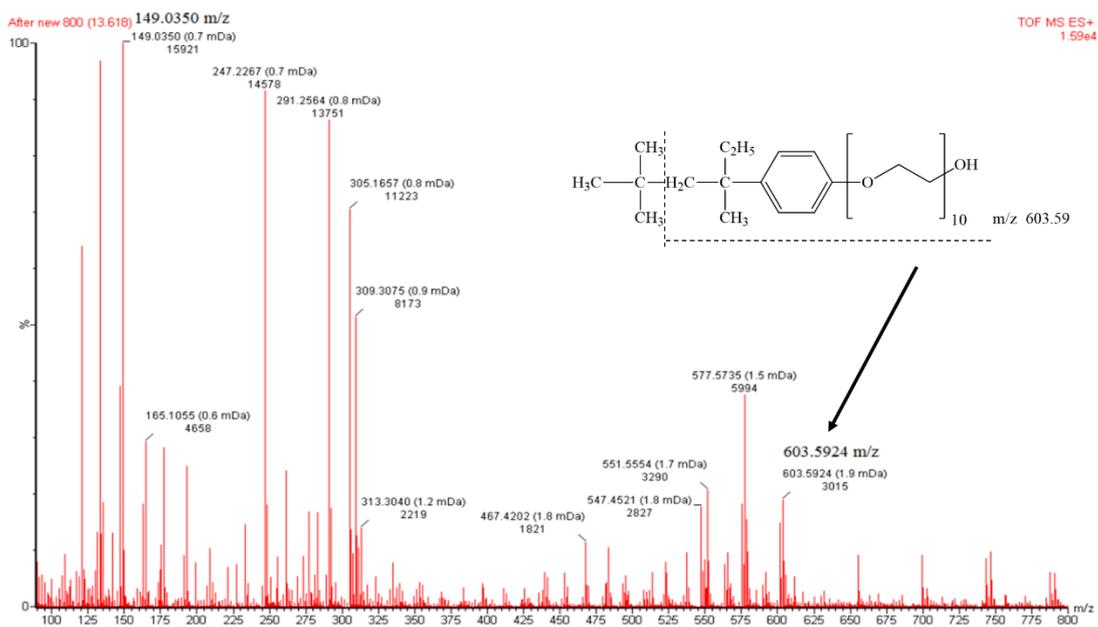

Figure 8. Ion chromatogram of degradation compound from NPE-10 at retention time of 13.61 min.

# List of table

Table 1. The percentage degradation of NPE-10 in acidic solution.

| Condition of NPE-10 degradation | Percentage of degradation (%) |
|---|---|
| Without 0.04 M Ce(III) | 61.44 |
| Containing 0.04 M Ce(III) | 85.93 |
| With 0.04 M Ce(III) and 0.009 M Ag(I) | 95.12 |